\documentclass{iau}

\usepackage[utf8]{inputenc}
\usepackage{tikz, graphicx, xcolor}
\usepackage{multirow}
\usepackage{natbib, aas_macros}

\newcommand{\Msun}{M_\odot}

\newcommand{\pc}{\mathrm{pc}}

\definecolor{lime}{HTML}{A6CE39}
\DeclareRobustCommand{\orcidicon}{%
	\begin{tikzpicture}
	\draw[lime, fill=lime] (0,0)
	circle [radius=0.16]
	node[white] {{\fontfamily{qag}\selectfont \tiny ID}};
	\draw[white, fill=white] (-0.0625,0.095)
	circle [radius=0.007];
	\end{tikzpicture}
	\hspace{-2mm}
}

\newcommand{\orcidVP}{\href{https://orcid.org/0000-0002-3031-062X}{\orcidicon}}
\newcommand{\orcidVK}{\href{https://orcid.org/0000-0002-5760-0459}{\orcidicon}}
\newcommand{\orcidBB}{\href{https://orcid.org/0000-0002-3578-6037}{\orcidicon}}
\newcommand{\orcidFP}{\href{https://orcid.org/0000-0002-9850-2708}{\orcidicon}}
\newcommand{\orcidAE}{\href{https://orcid.org/0000-0001-6049-3132}{\orcidicon}}

\begin{document}

\lefttitle{Pavlík et al.}
\righttitle{Does IRS13 require an intermediate-mass black hole?}

\jnlPage{1}{4}
\jnlDoiYr{2026}
\doival{10.1017/xxxxx}

\aopheadtitle{Proceedings IAU Symposium}
\editors{M. Zaja\v{c}ek,  T. Je\v{r}\'{a}bkov\'{a}, V. Karas, R. Schödel \&  P. Sukov\'{a}, eds.}

\title{Does IRS13 require\\[-5pt]an intermediate-mass black hole?}

\author{%
V\'aclav Pavl\'ik$^{1,2,\star}$\orcidVP,
Vladimír Karas$^{1}$\orcidVK,
Florian Pei{\ss}ker$^{3}$\orcidFP,
Bhavana Bhat \orcidBB, and
Andreas Eckart$^{3,4}$ \orcidAE
}

\affiliation{%
$^1$ Astronomical Institute, Czech Academy of Sciences, Bo\v{c}n\'i~II~1401, 141~00~Prague~4, Czech Republic\\
$^\star$ \texttt{pavlik@asu.cas.cz}\\
$^2$ Dept.~of Astronomy, Indiana University, Swain Hall West, 727 E 3rd St., Bloomington, IN 47405, USA\\
$^3$ I.~Physikalisches Institut der Universität zu Köln, Zülpicher Str.~77, D-50937 Köln, Germany\\
$^4$ Max Planck Institut für Radioastronomie, Auf dem Hügel 69, D-53121 Bonn, Germany}

\begin{abstract}
We investigate whether the Galactic-centre association IRS13 requires an intermediate-mass black hole (IMBH) to remain bound. Using high-precision $N$-body calculations of IRS13-like systems orbiting the Milky Way supermassive black hole (SMBH), as well as simulations of more massive infalling clusters, we find that an IRS13-sized cluster dissolves on a very short timescale even when an IMBH of $4\times10^4\,\Msun$ is present. The observed velocity dispersion can arise naturally from the tidal field of Sgr~A* and the infall event itself; therefore, it need not imply a central IMBH. In the infalling-cluster runs, tidal stripping produces transient disks, spiral-like structures, and ring-like overdensities. These features suggest a broader interpretation: IRS13 is best viewed as a temporary phase-space overdensity, analogous in mechanism (though not in scale) to phase-wrapped shells and streams in disrupted galaxies.
\end{abstract}


\maketitle

\section{Introduction}

The existence and growth of intermediate-mass black holes (IMBHs) remain open problems in stellar dynamics and black-hole formation. In the Galactic centre, IRS13 has long been discussed as a possible IMBH-hosting association \citep{maillard_etal2004, schodel_etal2005, fritz_etal2010, peissker_etal2023, peissker_etal2024}, but the dynamical evidence is still indirect. Our recent modelling results \citep{pavlik_IRS13} show that IRS13-like systems cannot be considered long-lived bound clusters at their distance from Sgr~A*. Consequently, the velocity dispersion of IRS13's members alone cannot be used to infer the presence or mass of a putative IMBH.

\section{Numerical models and results}

In the original study \citep{pavlik_IRS13}, we numerically follow the formation and dynamical evolution of IRS13-like systems. We use two approaches.

\subsection{Orbiting IRS13-like clusters}

\begin{figure}
    \centering
    \includegraphics[width=\linewidth]{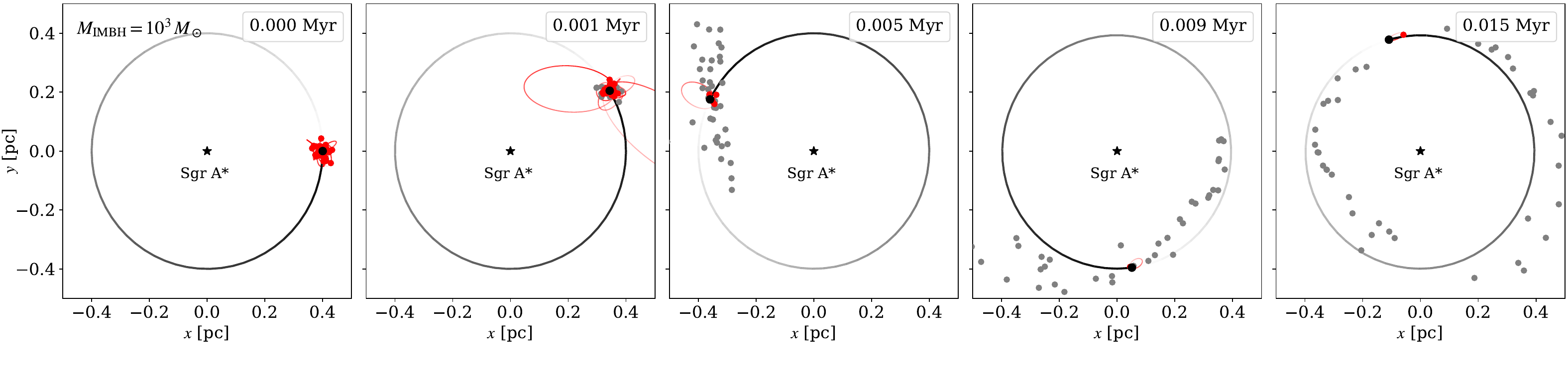}\\
    \includegraphics[width=\linewidth]{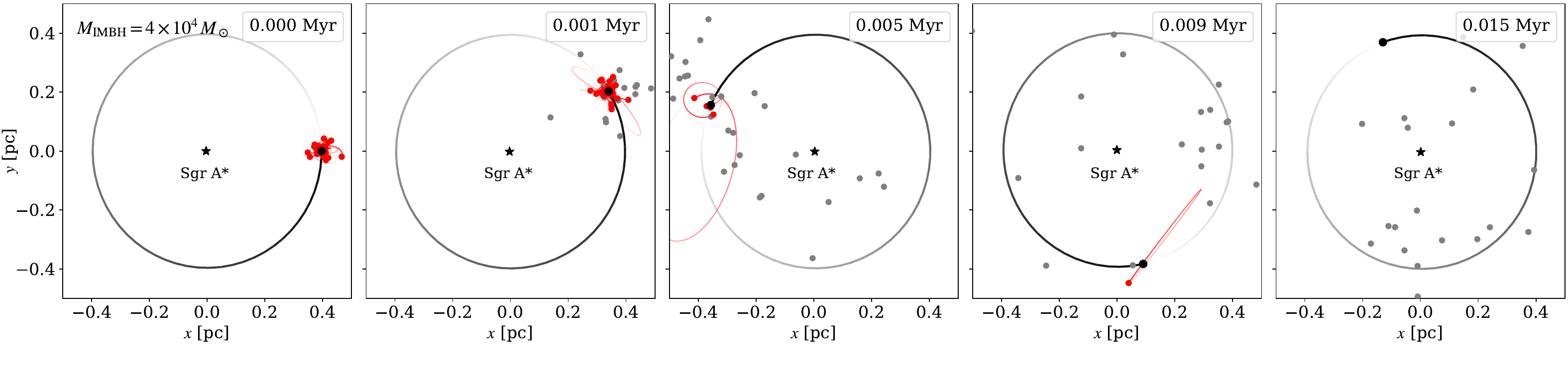}
    \caption{Snapshots illustrating the tidal disruption of representative IRS13-like models on initially circular orbits around Sgr~A* (as the black trajectory shows). The stars are initially bound to the IMBH (black circle) but rapidly disperse -- red points denote stars bound to the IMBH (with red curves showing their instantaneous osculating Keplerian orbits), whereas grey points represent unbound stars. The stellar mass is $1\,\Msun$, and the IMBH mass is indicated in each row.}
    \label{fig:orbit_circ}
\end{figure}

\begin{figure}
    \centering
    \vspace{20pt}
    \includegraphics[width=\linewidth]{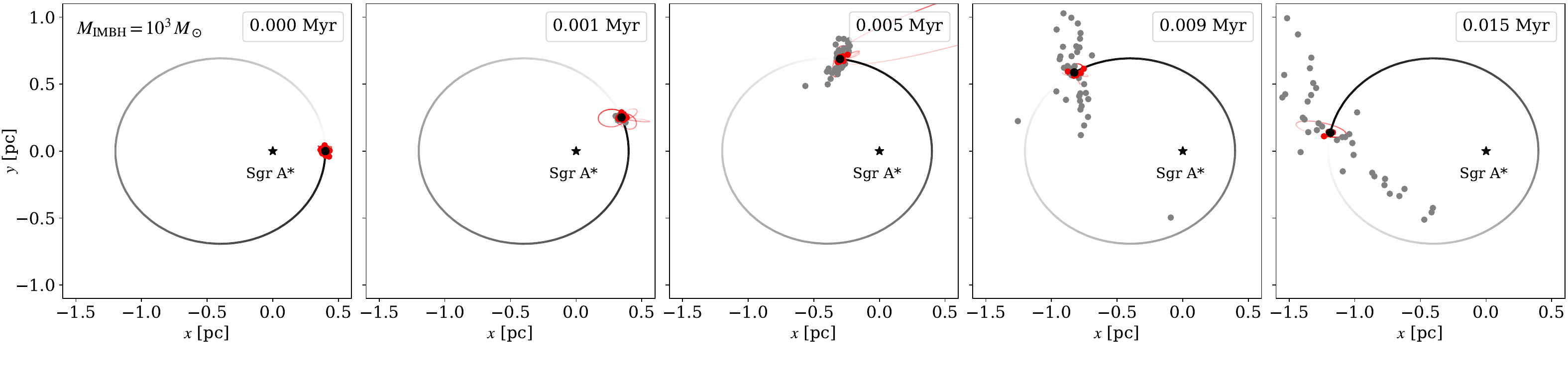}\\
    \includegraphics[width=\linewidth]{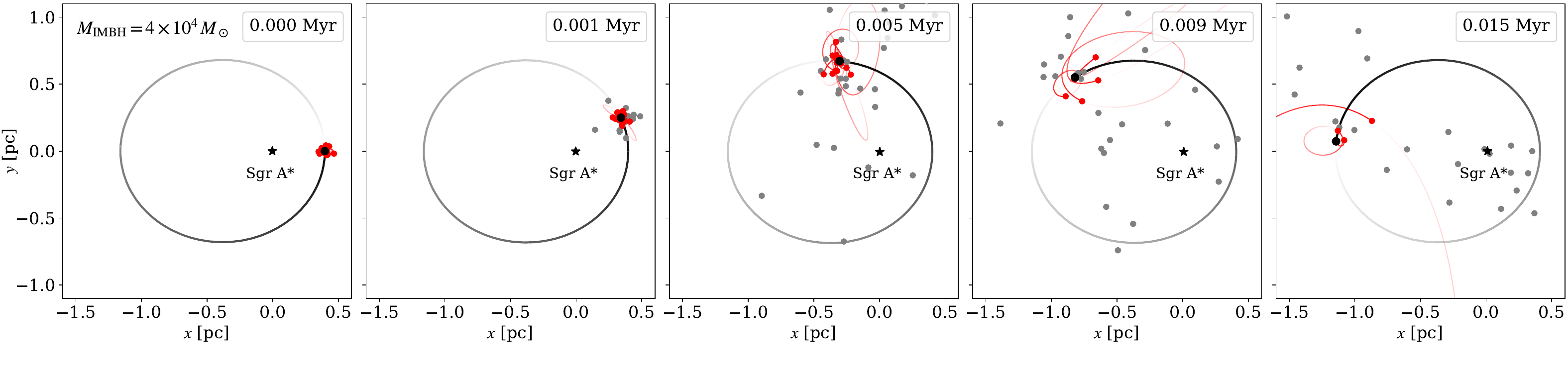}
    \caption{Same as Fig.~\ref{fig:orbit_circ} but for the initially eccentric orbit (as shown by the black trajectory).}
    \label{fig:orbit_ecc}
\end{figure}

First, we model IRS13-like clusters containing 44 stars \citep[which is the reported number of sources by][]{peissker_etal2023} in the tidal field of Sgr~A*, with and without an embedded IMBH. We vary the IMBH's mass, and we explore circular and eccentric orbits of the cluster. The systems are initially virialised, with equal-mass stellar populations (1 or $10\,M_\odot$), and are evolved with direct $N$-body methods \citep{rebound}.

We find that the orbiting-cluster models dissolve within a fraction of a Myr (see the illustrative cases in Figs.~\ref{fig:orbit_circ} and~\ref{fig:orbit_ecc}, or the full analysis of all models in \citealt{pavlik_IRS13}). Even the most massive IMBH considered \citep[$4\times10^4\,\Msun$, according to][]{peissker_etal2023} cannot prevent disruption.\!\footnote{However, the existence of such a massive IMBH so close to Sgr~A* has been excluded, e.g., by \citet{gualandris_merritt2009} and \citet{rb2004}.} The dissolution proceeds somewhat more slowly on eccentric orbits, where the cluster spends most of its time farther from Sgr~A* and is therefore exposed to a weaker tidal field.

Conversely, increasing the IMBH mass accelerates the disruption through stronger dynamical interactions between the stars and the IMBH. In all models, a hard star--IMBH binary forms shortly after the start of the simulation. It causes dynamical heating of the cluster and efficient ejection of stars despite the larger Hill sphere of a more massive IMBH (compare the compactness of the clusters and the number of bound stars between the top and the bottom row within Figs.~\ref{fig:orbit_circ} and~\ref{fig:orbit_ecc}). The stellar escape is therefore driven by the combined action of the SMBH tidal field and IMBH-induced dynamical heating.

\subsection{Infalling clusters}

\begin{figure}
    \centering
    \includegraphics[width=\linewidth]{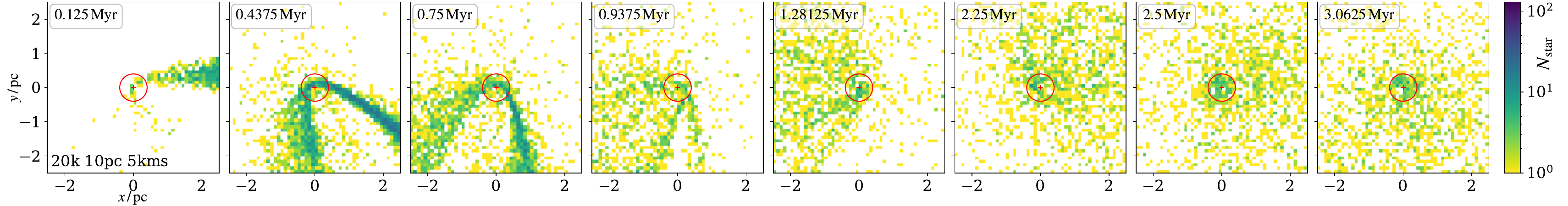}\\
    \includegraphics[width=\linewidth]{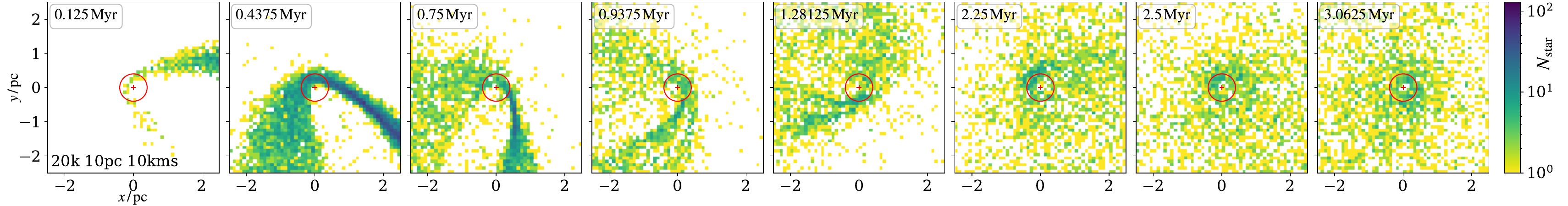}\\
    \includegraphics[width=\linewidth]{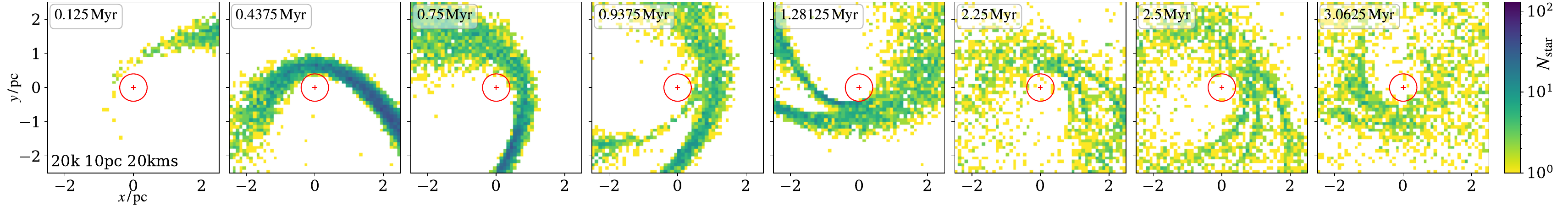}\\
    \includegraphics[width=\linewidth]{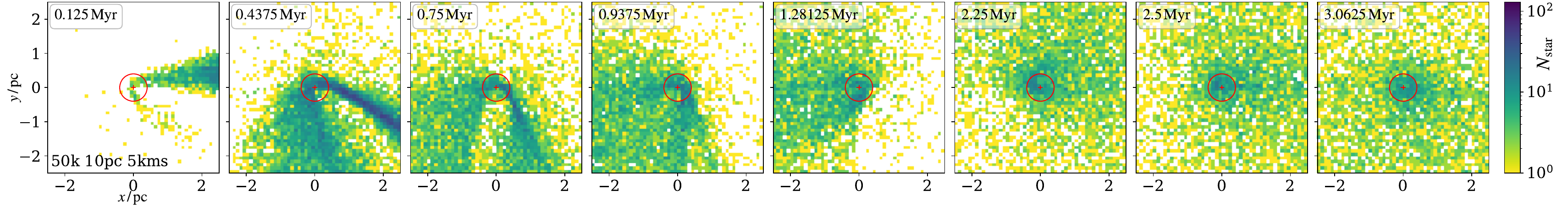}\\
    \includegraphics[width=\linewidth]{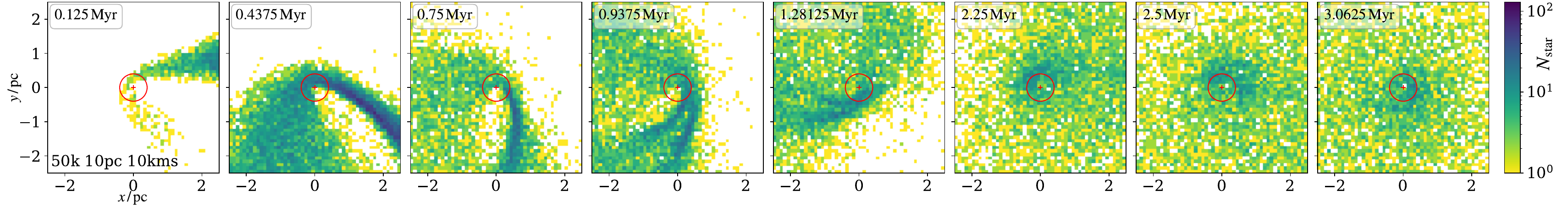}
    \caption{Illustrative snapshots of five simulated infalling cluster models (the initial number of stars, distance from Sgr~A* in the $x$ direction, and tangential velocity are indicated in the corner of each row). The colour scale shows the number of stars in each bin. The red cross marks the position of Sgr~A* (orbited by the IMBH at $0.4\,\pc$, see the red circle). Tidal stripping during successive passages through the Galactic centre produces transient spiral-like and ring-like overdensities, which gradually disperse.}
    \label{fig:infall}
\end{figure}

To place IRS13 into a broader formation context, we performed a second suite of direct $N$-body simulations of more massive star clusters initially located at galactocentric distances of ${\sim}10$--$100\,\pc$. The clusters were evolved in the external Galactic potential, including Sgr~A* and a putative IMBH on a circular orbit, using the \textsc{PeTar} code \citep{petar}. We explored a range of cluster masses, sizes, and orbital parameters to follow their tidal disruption during inspiral towards the Galactic centre.

As the clusters lose mass through repeated pericentre passages, they become progressively disrupted by the tidal field of the SMBH. The stripped stars form transient disks, spiral-like streams, expanding ring-like overdensities, and temporary stellar clumps before they gradually disperse and phase-mix (see Fig.~\ref{fig:infall}). Depending on the initial orbit and compactness of the cluster, a significant fraction of the stars can be deposited into the central parsec and naturally produce IRS13-like stellar overdensities without requiring them to remain gravitationally bound \citep[see the full analysis in][]{pavlik_IRS13}.

\section{Discussion}

Our simulations suggest that IRS13 is more likely a transient overdensity formed during the tidal disruption of an infalling star cluster than a long-lived bound system. This scenario also places IRS13 into a broader astrophysical context. Although the resulting structures differ morphologically from the shells and ripples observed on a larger scale in interacting galaxies, they originate from the same physical mechanism \citep[cf.][]{1984ApJ...279..596Q,1986A&A...166...53D,2016A&A...588A..77B}. In both cases, tidal stripping of a collisionless stellar system is followed by phase wrapping and phase mixing in a smooth gravitational potential.

This interpretation also has important consequences for the dynamical analysis of IRS13. The association is neither isolated nor in dynamical equilibrium, and its stellar population changes continuously. Consequently, the observed velocity dispersion cannot be interpreted using standard equilibrium dynamical methods and does not provide a robust dynamical constraint on the presence or mass of a putative IMBH.

An observational test of this scenario would be to search for coherent phase-space substructures and tidal features around IRS13 and Sgr~A* that may have formed during the disruption. Such features would also connect the evolution of infalling clusters on scales of tens of parsecs with transient stellar structures in the central parsec.

\begin{acknowledgements}
VP has received funding from the European Union's Horizon Europe and the Central Bohemian Region under the Marie Skłodowska-Curie Actions -- COFUND (\href{https://doi.org/10.3030/101081195}{ID~101081195} ``MERIT'').
This study used computational resources provided by the e-INFRA CZ project (ID:90254), supported by the Ministry of Education, Youth and Sports of the Czech Republic; and the computational cluster VIRGO at the Astronomical Institute of the Czech Academy of Sciences.
VP and VK also acknowledge the support from the project RVO:67985815 at the Czech Academy of Sciences.
\end{acknowledgements}

\bibliographystyle{iaulike}
\bibliography{main}

@article{petar,
archivePrefix = {arXiv},
arxivId = {2006.16560},
author = {Wang, Long and Iwasawa, Masaki and Nitadori, Keigo and Makino, Junichiro},
doi = {10.1093/mnras/staa1915},
eprint = {2006.16560},
issn = {13652966},
journal = {\mnras},
month = {6},
number = {1},
pages = {536--555},
publisher = {Oxford University Press},
title = {{Petar: A high-performance N-body code for modelling massive collisional stellar systems}},
url = {https://arxiv.org/abs/2006.16560v2},
volume = {497},
year = {2020}
}

@ARTICLE{peissker_etal2023,
       author = {{Pei{\ss}ker}, Florian and {Zaja{\v{c}}ek}, Michal and {Thomkins}, Lauritz and {Eckart}, Andreas and {Labadie}, Lucas and {Karas}, Vladim{\'\i}r and {Sabha}, Nadeen B. and {Steiniger}, Lukas and {Melamed}, Maria},
        title = "{The Evaporating Massive Embedded Stellar Cluster IRS 13 Close to Sgr A*. I. Detection of a Rich Population of Dusty Objects in the IRS 13 Cluster}",
      journal = {\apj},
         year = 2023,
        month = oct,
       volume = {956},
       number = {2},
          eid = {70},
        pages = {70},
          doi = {10.3847/1538-4357/acf6b5},
archivePrefix = {arXiv},
       eprint = {2310.06156},
 primaryClass = {astro-ph.GA},
       adsurl = {https://ui.adsabs.harvard.edu/abs/2023ApJ...956...70P}
}

@ARTICLE{peissker_etal2024,
       author = {{Pei{\ss}ker}, Florian and {Zaja{\v{c}}ek}, Michal and {Labaj}, Mat{\'u}{\v{s}} and {Thomkins}, Lauritz and {Elbe}, Andreas and {Eckart}, Andreas and {Labadie}, Lucas and {Karas}, Vladim{\'\i}r and {Sabha}, Nadeen B. and {Steiniger}, Lukas and {Melamed}, Maria},
        title = "{The Evaporating Massive Embedded Stellar Cluster IRS 13 Close to Sgr A*. II. Kinematic Structure}",
      journal = {\apj},
         year = 2024,
        month = jul,
       volume = {970},
       number = {1},
          eid = {74},
        pages = {74},
          doi = {10.3847/1538-4357/ad4098},
archivePrefix = {arXiv},
       eprint = {2407.15800},
 primaryClass = {astro-ph.GA},
       adsurl = {https://ui.adsabs.harvard.edu/abs/2024ApJ...970...74P}
}

@ARTICLE{maillard_etal2004,
       author = {{Maillard}, J.~P. and {Paumard}, T. and {Stolovy}, S.~R. and {Rigaut}, F.},
        title = "{The nature of the Galactic Center source IRS 13 revealed by high spatial resolution in the infrared}",
      journal = {\aap},
         year = 2004,
        month = aug,
       volume = {423},
        pages = {155-167},
          doi = {10.1051/0004-6361:20034147},
archivePrefix = {arXiv},
       eprint = {astro-ph/0404450},
 primaryClass = {astro-ph},
       adsurl = {https://ui.adsabs.harvard.edu/abs/2004A&A...423..155M}
}

@ARTICLE{schodel_etal2005,
       author = {{Sch{\"o}del}, R. and {Eckart}, A. and {Iserlohe}, C. and {Genzel}, R. and {Ott}, T.},
        title = "{A Black Hole in the Galactic Center Complex IRS 13E?}",
      journal = {\apjl},
         year = 2005,
        month = jun,
       volume = {625},
       number = {2},
        pages = {L111-L114},
          doi = {10.1086/431307},
archivePrefix = {arXiv},
       eprint = {astro-ph/0504474},
 primaryClass = {astro-ph},
       adsurl = {https://ui.adsabs.harvard.edu/abs/2005ApJ...625L.111S}
}

@ARTICLE{fritz_etal2010,
       author = {{Fritz}, T.~K. and {Gillessen}, S. and {Dodds-Eden}, K. and {Martins}, F. and {Bartko}, H. and {Genzel}, R. and {Paumard}, T. and {Ott}, T. and {Pfuhl}, O. and {Trippe}, S. and {Eisenhauer}, F. and {Gratadour}, D.},
        title = "{GC-IRS13E{\textemdash}A Puzzling Association of Three Early-type Stars}",
      journal = {\apj},
         year = 2010,
        month = sep,
       volume = {721},
       number = {1},
        pages = {395-411},
          doi = {10.1088/0004-637X/721/1/395},
archivePrefix = {arXiv},
       eprint = {1003.1717},
 primaryClass = {astro-ph.GA},
       adsurl = {https://ui.adsabs.harvard.edu/abs/2010ApJ...721..395F}
}

@ARTICLE{rebound,
       author = {{Rein}, H. and {Liu}, S. -F.},
        title = "{REBOUND: an open-source multi-purpose N-body code for collisional dynamics}",
      journal = {\aap},
         year = 2012,
        month = jan,
       volume = {537},
          eid = {A128},
        pages = {A128},
          doi = {10.1051/0004-6361/201118085},
archivePrefix = {arXiv},
       eprint = {1110.4876},
 primaryClass = {astro-ph.EP},
       adsurl = {https://ui.adsabs.harvard.edu/abs/2012A&A...537A.128R}
}

@ARTICLE{gualandris_merritt2009,
       author = {{Gualandris}, Alessia and {Merritt}, David},
        title = "{Perturbations of Intermediate-mass Black Holes on Stellar Orbits in the Galactic Center}",
      journal = {\apj},
         year = 2009,
        month = nov,
       volume = {705},
       number = {1},
        pages = {361-371},
          doi = {10.1088/0004-637X/705/1/361},
archivePrefix = {arXiv},
       eprint = {0905.4514},
 primaryClass = {astro-ph.GA},
       adsurl = {https://ui.adsabs.harvard.edu/abs/2009ApJ...705..361G}
}

@ARTICLE{rb2004,
       author = {{Reid}, M.~J. and {Brunthaler}, A.},
        title = "{The Proper Motion of Sagittarius A*. II. The Mass of Sagittarius A*}",
      journal = {\apj},
         year = 2004,
        month = dec,
       volume = {616},
       number = {2},
        pages = {872-884},
          doi = {10.1086/424960},
archivePrefix = {arXiv},
       eprint = {astro-ph/0408107},
 primaryClass = {astro-ph},
       adsurl = {https://ui.adsabs.harvard.edu/abs/2004ApJ...616..872R}
}

@ARTICLE{pavlik_IRS13,
       author = {{Pavl{\'\i}k}, V{\'a}clav and {Karas}, Vladim{\'\i}r and {Bhat}, Bhavana and {Pei{\ss}ker}, Florian and {Eckart}, Andreas},
        title = "{Dynamics of star associations in an SMBH--IMBH system: The case of IRS13 in the Galactic centre}",
      journal = {\aap},
         year = 2024,
        month = dec,
       volume = {692},
          eid = {A104},
        pages = {A104},
          doi = {10.1051/0004-6361/202452050},
archivePrefix = {arXiv},
       eprint = {2410.14541},
 primaryClass = {astro-ph.GA},
       adsurl = {https://ui.adsabs.harvard.edu/abs/2024A&A...692A.104P}
}

@ARTICLE{2016A&A...588A..77B,
       author = {{B{\'\i}lek}, M. and {Cuillandre}, J.-C. and {Gwyn}, S. and {Ebrov{\'a}}, I. and {Barto{\v{s}}kov{\'a}}, K. and {Jungwiert}, B. and {J{\'\i}lkov{\'a}}, L.},
        title = "{Deep imaging of the shell elliptical galaxy NGC 3923 with MegaCam}",
      journal = {\aap},
         year = 2016,
        month = apr,
       volume = {588},
          eid = {A77},
        pages = {A77},
          doi = {10.1051/0004-6361/201526608},
archivePrefix = {arXiv},
       eprint = {1505.07146},
 primaryClass = {astro-ph.GA},
       adsurl = {https://ui.adsabs.harvard.edu/abs/2016A&A...588A..77B}
}

@ARTICLE{1986A&A...166...53D,
       author = {{Dupraz}, C. and {Combes}, F.},
        title = "{Shells around galaxies : testing the mass distribution and the 3-D shape of ellipticals.}",
      journal = {\aap},
         year = 1986,
        month = sep,
       volume = {166},
        pages = {53-74},
       adsurl = {https://ui.adsabs.harvard.edu/abs/1986A&A...166...53D}
}

@ARTICLE{1984ApJ...279..596Q,
       author = {{Quinn}, P.~J.},
        title = "{On the formation and dynamics of shells around elliptical galaxies.}",
      journal = {\apj},
         year = 1984,
        month = apr,
       volume = {279},
        pages = {596-609},
          doi = {10.1086/161924},
       adsurl = {https://ui.adsabs.harvard.edu/abs/1984ApJ...279..596Q}
}

\end{document}